\documentclass[a4paper,11pt]{article}

\usepackage{jcappub} 

\usepackage[T1]{fontenc} 
\usepackage{amsmath, amssymb, amsfonts,mathrsfs,graphicx,graphics}
\usepackage{dsfont}
\usepackage{indentfirst}

\newcommand {\bb}[1]{\mbox{\boldmath $#1$}}
\newcommand{\ii}{\mathrm{i}}
\newcommand{\Tr}{\mathrm{Tr}}

\title{\boldmath {Spin rotation of neutrinos produced by compact magnetized
	astrophysical objects}}

\author{A. V. Chukhnova}
\author{and A. E. Lobanov}

\affiliation{Department of Theoretical Physics, Faculty of Physics,
	Moscow State University,\\ 119991 Moscow, Russia}

\emailAdd{av.chukhnova@physics.msu.ru}
\emailAdd{lobanov@phys.msu.ru}

\abstract{We study the propagation of neutrinos from compact astrophysical objects with strong magnetic field, such as magnetars. Both neutrino spin rotation and oscillations in a realistic three-flavor model are taken into account. We solve the neutrino evolution equation in the magnetic field near the object and obtain the probabilities of all possible spin-flavor transitions.  Assuming that neutrinos are produced in the interior of the star being in their left-handed state with electron flavor, we use these spin-flavor transition probabilities to estimate the fraction of neutrinos, which change their helicity during propagation to the Earth and thus become unobservable.  Using the model of the magnetar with dipolar magnetic field, we demonstrate anisotropy in the left-handed and right-handed neutrino fluxes from such astrophysical objects.}

\begin{document}

\maketitle

\flushbottom

\section*{Introduction}

Neutrino astrophysics is an important area of study, which can give us an insight into the depths of the Universe, since the neutrino fluxes can travel at long distances almost without being destroyed, reflected or defracted. However, the neutrino state does not remain the same during propagation from the production point to the Earth, and so the neutrino may be detected in a state different from its initial one. That is why we need an accurate and precise description of the neutrino evolution both in compact astrophysical objects and in the surrounding space in order to understand the processes inside the stars.

The first thing, which should certainly be taken into account when we describe neutrino propagation, is the phenomenon of neutrino oscillations. The phenomenological theory of oscillations, based on the ideas by Pontecorvo \cite{Pontecorvo1957} (for more detail see, e.g. \cite{Bilenky1978}), is suitable for the propagation of high-energy neutrinos in vacuum or in matter at rest \cite{Wolfenstein1978,MS_en}. However, in the general case even for ultrarelativistic neutrinos one should take into account the motion and the polarization of the background medium and the direct interaction of neutrino magnetic moments with the background field. For this purpose the theory of neutrino propagation should be modified in a more accurate way, which is more reasonable in the mathematical sense.

For neutrinos propagating in moving or polarized matter and magnetic field it is not enough to study the evolution of neutrino flavor as it can be done for neutrino in matter at rest. When the neutrino propagates in moving or polarized matter \cite{Lobanov2001}, or when electromagnetic field is present \cite{Fujikawa1980,Shrock1982,Okun1986,Voloshin1986}, the neutrino helicity also does not remain the same. Since the neutrino interaction with matter and electromagnetic field depends significantly on neutrino polarization, it is necessary to take into account the spin rotation effect and the neutrino oscillations simultaneously. Note, that the effect of neutrino spin rotation has been studied for a long time \cite{Cisneros1971,Akhmedov1988b,Semikoz1987_en,Semikoz1989_en,LiM,Pantaleone1992,
Dobrynina2016,Vlasenko2014,Kartavtsev2015,Ternov2019}. In particular, in \cite{PR2020} we have derived a wave equation, which provides the opportunity to study neutrino flavor oscillations and spin rotation simultaneously in matter and electromagnetic field, and present spin-flavor transition probabilities for the case of constant external conditions.

The neutrino interaction with electromagnetic field may be significant only when the values of the fields are extremely high. The values of the magnetic induction up to $10^{14}-10^{15}$ G can be reached at the surface of magnetars (see, e.g., \cite{Turolla_2015,Beloborodov_2017}), and perhaps even higher values in the core. The information about the fields magnetars can be found in the online McGill magnetar catalogue\cite{McGill_catague}. In \cite{PR2020} we have demonstrated that the effect of spin rotation in magnetic field can hardly be observed in presently known astrophysical objects, since extremely high electromagnetic field is required in a huge area of space.
Let us consider a Dirac neutrino with the mass about $m_\nu=0.033 $ eV and the magnetic moment $\mu_\nu$ given by the Standard Model predictions.
Even for the constant magnetic field $10^{16}$ G the characteristic length of neutrino spin rotation is $L = \pi/ (\mu_\nu B)\approx 1070$ km,
which exceeds the typical magnetar radius significantly. However, since we actually do not know the values of the neutrino masses and the magnetic moments, the maximal possible values of fields, generated by compact astrophysical objects, and also the exact size of such objects, this estimation does not prove that observing the spin rotation effect is completely impossible. The spin rotation effect will be caused by astrophysical objects with realistic size, if in this objects the magnetic field is much higher than $10^{16}$ G.  Here we also call such objects ``magnetars'' by analogy to presently known astrophysical objects. In this paper we investigate the neutrino propagation in hypothetical magnetars with higher fields or larger size than the currently known ones or in the case of larger neutrino magnetic moments than those predicted by the Standard Model.
Note, that the present experimental limit on the neutrino magnetic moment is $\mu_\nu<2.9\cdot 10^{-11} \mu_B$ \cite{Agostini2017, Beda2013}, and for the chosen value of neutrino mass $m_\nu$ that is about nine orders of magnitude higher, than the Standard Model theoretical prediction.

In Sec. \ref{sec1} we present the formalism used in our research.
In Sec. \ref{sec2} we obtain the spin-flavor transition probabilities for the neutrinos from compact astrophysical objects. For this purpose we assume that the neutrino is generated in the inner part of the magnetar in the left-handed state with electron flavor and demonstrate that after its propagation through the dense regions of the magnetar this state remains almost the same. The influence of these assumptions on the final results is discussed in Conclusion.
The details of our calculations can be found in Appendix. We suppose that the distance between the neutrino source and the detection point is high enough for the magnetic field to decrease completely, since we are mostly interested in the case, when the neutrino could be detected in terrestrial conditions.
In Sec. \ref{sec4} we study a model of the magnetar with dipolar magnetic field around it. As a result of our research we reveal that the spin-flip probability depends significantly on the angle between the neutrino velocity and the magnetic axis, that means, we show that the magnetars can generate anisotropic fluxes of right-handed and left-handed neutrinos.

\section{Neutrino evolution in matter and electromagnetic field}\label{sec1}
We study both spin rotation and flavor oscillations of the neutrino propagating in matter and electromagnetic field in the framework of the modification of the Standard Model, which was introduced in \cite{lobanov2019,tmf2017_en}. This modification of the Standard Model reproduces the well-known results of the phenomenological approach in the range of neutrino energies, where the phenomenological theory is valid, and requires no new types of particles or interactions. In our model both the neutrino oscillations and spin rotation are described in the framework of the quantum field theoretical approach in a mathematically rigorous way. This fact enables us to study some effects, which it is rather hard to formalize within the phenomenological approach. In particular, in \cite{PR2020,adia_arxiv} we discover the resonance behavior of spin-flavor transition probabilities due to neutrino transition magnetic moments.

Within our model the neutrinos with different masses are combined in $SU(3)$-multiplets. A $12$-component neutrino wave function determines the state of the whole multiplet, i.e. the mass structure of the multiplet and the spin states of its components.
Since in this model any superposition of the neutrino mass states is a pure quantum state of the neutrino multiplet, all the methods of quantum field theory are applicable. We take into account the interaction with electromagnetic field and matter composed of electrons, protons and neutrons using the modified Dirac equation for the neutrino multiplet \cite{PR2020}. In the quasi-classical approximation, which is a very good approximation for ultra-relativistic particles, the neutrino can be considered as moving with the constant $4$-velocity $u^\mu$. Then the neutrino wave equation takes the form
\begin{equation}\label{QQ}
\begin{array}{l}\displaystyle
\left(\mathrm{i} \frac{d}{d \tau}
\mathds{I}-\mathds{M}-\frac{1}{2}\left(f^{(e)} u\right)
\mathds{P}^{(e)}-\frac{1}{2}\left(f^{(\mathrm{N})} u\right) \mathds{I}-\frac{1}{2}
R_{e} \mathds{P}^{(e)} \gamma^{5} \gamma^{\sigma} s_{\sigma}^{(e)} \gamma^{\mu}
u_{\mu} \right. \\ \left.  \displaystyle
-\frac{1}{2} R_{\mathrm{N}} \mathds{I} \gamma^{5} \gamma^{\sigma}
s_{\sigma}^{(\mathrm{N})} \gamma^{\mu} u_{\mu}-\mu_{0} \mathds{M} \gamma^{5}
\gamma^{\mu *} F_{\mu \nu} u^{\nu}-\mathds{M}_{h} \gamma^{5} \gamma^{\mu \star}
F_{\mu \nu} u^{\nu}
-\mathds{M}_{a h} \gamma^{5} \gamma^{\mu} F_{\mu \nu} u^{\nu} \right)
\varPsi(\tau) =0
\end{array}
\end{equation}
\noindent where $\tau$ is the neutrino proper time, and $\varPsi(\tau)$ is a $12$-component quasi-classical neutrino wave function, which satisfies the relation $\gamma^\mu u_\mu \varPsi(\tau) = \varPsi(\tau)$. The proper time is related to the neutrino path length $L$ by the relation $\tau = L/{|{\bf u}|}$. In Eq. \eqref{QQ} $\mathds{I}$ is $3\times3$ identity matrix, $\mathds{M}$ is	the neutrino mass matrix,  $\mathds{P}^{(e)}$ is the projection operator on the
neutrino state with electron flavor, $F^{\mu\nu}$ is the electromagnetic field tensor,	 ${}^{\star\!}F^{\mu\nu} =
-\frac{1}{2}e^{\mu\nu\rho\lambda}F_{\rho\lambda}$  is the dual electromagnetic field tensor. The interaction with the neutrino transition magnetic moments \cite{Shrock1982} (see also \cite{Giunti2015}) is taken into account with the help of the Hermitian matrices $\mathds{M}_h$ and $\mathds{M}_{ah}$. The diagonal magnetic moments of the Standard Model neutrinos are in the first approximation	proportional to the neutrino masses, i.e. the matrix of the diagonal magnetic moments is defined as $\mu_0 \mathds{M}$, where
\begin{equation}
\mu_0 = \frac{3 e \mathrm{G_F}}{8\sqrt{2}\pi^2}.
\end{equation}

The interaction with matter is taken into account using two effective potentials $f^{\alpha (\mathrm e)}$ and $f^{\alpha (\mathrm N)}$, which generalize the potential used in \cite{Wolfenstein1978}.
These potentials are associated with the currents $j^{\alpha{(i)}}$ and polarizations $\lambda^{\alpha{(i)}}$ of the background fermions of the type $(i)$, $i=e,p,n$ for electrons, protons and neutrons, correspondingly
\begin{equation}\label{l10}
j^{\alpha{(i)}}=\{\bar{n}^{(i)} v^{0{(i)}},\bar{n}^{(i)}{\bf{v}}^{{(i)}}\},
\end{equation}
\begin{equation}\label{l11}
\lambda^{\alpha{(i)}} =
\left\{\bar{n}^{(i)}
({\bb{\zeta}}^{(i)}{\bf{v}}^{{(i)}}), \bar{n}^{(i)}\left({\bb{\zeta}}^{(i)} + \frac{
	{\bf{v}}^{{(i)}} ({\bb{\zeta}}^{(i)}{\bf{v}}^{{(i)}})}
{1+v^{0{(i)}}}\right)\right\}.
\end{equation}
\noindent These 4-current and 4-polarization vectors characterize the medium as a whole. In these formulas
$\bar{n}^{(i)}$ and ${\bb{\zeta}}^{(i)}
\;(0\leqslant |{\bb{\zeta}}^{(i)} |^2 \leqslant 1)$
are the number densities and the average value of the polarization vector of the background
fermions in the reference frame in which the
average momentum of fermions $(i)$ is equal to zero. The $4$-velocity of this reference frame is
denoted as $v^{\alpha{(i)}}
=\{v^{0{(i)}},{\bf{v}}^{{(i)}}\}$ .

The potential
\begin{equation}\label{l12x}
f^{\alpha{(e)}} =\sqrt{2}{G}_{{\mathrm F}}\left({j^{\alpha{(e)}}}
-\lambda^{\alpha{(e)}}\right)
\end{equation}
\noindent determines the neutrino interaction with electrons via the charged currents, while the potential
\begin{equation}\label{l14x}
f^{\alpha (\mathrm N)} =\sqrt{2}{G}_{{\mathrm F}}\sum\limits_{i}
\left({j^{\alpha{(i)}}}
\left(T^{(i)}-2Q^{(i)}\sin^{2}
\theta_{\mathrm{W}}\right)-
{\lambda^{\alpha{(i)}}}T^{(i)}\right)
\end{equation}
\noindent determines the interaction via neutral currents.  Here  $\theta_\mathrm{W}$ is the Weinberg angle, $T^{(i)}$  is the
weak isospin projection of the background fermion $(i)$,  $Q^{(i)}$ is the
electric charge of the background fermion in the units of the positron charge.

\noindent In Eq. \eqref{QQ} we use the following notations
\begin{equation}\label{t4}
{R}(f)={\sqrt{(fu)^2 - f^2}}, \quad \displaystyle
s_{\mu}(f)=\frac{u_{\mu}(fu)-f_{\mu}}{\sqrt{(fu)^2-f^2}},
\end{equation}
\begin{equation}\label{t5}
\begin{array}{l} \displaystyle
{R}_e={R}(f^{(e)}), \qquad {R}_\mathrm{N}={R}(f^{(\mathrm{N})}), \qquad
s_\mu^{(e)} = s_\mu(f^{(e)}), \qquad s_\mu^{(\mathrm{N})} = s_\mu (f^{(\mathrm{N})}).
\end{array}
\end{equation}

Since within this approach the neutrino wave function describes both the mass structure of the neutrino multiplet and its spin state, to obtain the information about final neutrino flavor and helicity, we need only to solve the neutrino evolution equation, i.e. to write the neutrino multiplet wave function. Then all the transition probabilities can be derived using quantum mechanical formulas. In the case of arbitrary external conditions this procedure obviously can not be performed analytically. For homogeneous fields and matter with constant velocity and polarization the general expression for the spin-flavor transition probabilities is presented in \cite{PR2020}.

\section{Neutrino evolution in the magnetosphere}\label{sec2}

To obtain the probabilities of neutrino transitions between the states with definite helicity and flavor, one needs to solve the neutrino evolution equation \eqref{QQ}. However, since the density profile, the structure of the matter fluxes and the configuration of the magnetic field inside the magnetar are not clear,
we study analytically  the propagation of the neutrino outside it, assuming that the field decreases with the distance.

We suppose that at the edge of the magnetar the neutrino has left-handed helicity and electron flavor.
In our opinion this assumption is very realistic for the following reasons.
The matter of the magnetar can be considered to be at rest as a whole. Then the matter current and
the polarization in Eq. \eqref{QQ} take the form
\begin{equation}\label{5}
j^{\alpha{(i)}}=\{n^{(i)}, {\bf 0}\},\quad \lambda^{\alpha{(i)}}=\{0,n^{(i)}
{\bb \zeta}^{(i)}\},
\end{equation}
\noindent where $n^{(i)}$ is the number density of the fermion $(i)$.
Neutrino evolution equation \eqref{QQ} is derived for the medium, which consists of protons, neutrons and electrons. However, even if the core of the magnetar consists of quarks-gluon plasma rather than of neutrons and protons, Eq. \eqref{QQ} is also applicable. In this case to obtain the potential $f^{\alpha (\mathrm{N})}$  one needs to take the sum over the fermions $i=e,u,d$ (electrons, $u$-quarks and $d$-quarks) in Eq. \eqref{l14x}, since the amount of heavy quarks in the core of the magnetar can be expected to be rather low. Thus, no differences arise, and all the results obtained in this section for proton-neutron medium are still valid for quark-gluon plasma.

The  matter density of the magnetar can be about  $n\sim 10^{34}-10^{35}$ $\mathrm{sm}^{-3}$ and even reach the values up to $10^{40}$ $\mathrm{sm}^{-3}$ in the center. Taking into account the Standard model neutrino interaction, we expect that the neutrinos in the interior of the magnetar are produced  as left-handed neutrinos with the electron flavor. When the neutrino propagates inside the magnetar, the flavor oscillations of neutrinos with electron flavor are strongly suppressed due to extremely large values of the Wolfenstein potential \cite{Wolfenstein1978}.

For neutrino spin rotation both charged current and neutral current interactions are to be taken into account (see, e.g., \cite{arlomur}). Since it can be expected that for all fermions of the medium $|{\bb \zeta}^{(i)}|\ll 1$, the main role play the zero components of the effective potentials $f^{\alpha (e)}$ and $f^{\alpha (\mathrm{N})}$.
The influence of the magnetic field due to the direct interaction with the neutrino anomalous magnetic moments can be neglected when
\begin{equation}\label{6z}
\sqrt{2}G_{{\mathrm F}}\Big|n^{(e)}-\frac{1}{2} n^{(n)}\Big| \gg \frac{3eG_{{\mathrm F}}Bm_{\nu}}{8\sqrt{2}\pi^{2}},
\end{equation}
\noindent that is when the energies of the neutrino interaction with the field is much less than the energy of the neutrino interaction with the medium. Here we present the corresponding estimation in the Gauss units to make it more easy to obtain the numerical value of the field. That is, for the interior of the magnetar the following values are required for the magnetic field to become significant
\begin{equation}\label{6x}
\frac{B}{B_{0}}\gtrsim
\frac{16}{3}\pi^{2}\frac{m_{e}}{m_{\nu}}\left(\frac{\hbar}{m_{e}c}\right)^{3}\left|n^{(e)}-\frac{1}{2}n^{(n)}\right|,
\end{equation}
\noindent where $B_{0}\approx 4,41\cdot 10^{13}$ G is the Schwinger magnetic field. If we assume $m_{\nu}=0.033$ eV, then the expression \eqref{6x} takes the form
\begin{equation}\label{6xx}
B\gtrsim 1,5\cdot10^{-23}\,B_0 \cdot \left|n^{(e)} - \frac{1}{2}n^{(n)}\right|\, (\text{sm}^{-3}).
\end{equation}
The amount of electrons in the center of the star seams to be rather low and we can expect that $n^{(e)}-\frac{1}{2}n^{(n)}<0$. If the shell of the magnetar is composed of iron, then the sign of $n^{(e)}~-~\frac{1}{2}n^{(n)}$ changes somewhere inside the magnetar. Therefore, a part of the neutrino trajectory may lie in the region, where the relation \eqref{6xx} holds, and the influence of the magnetic field can cause spin rotation effect, if the density varies slow enough.
The possibility of such phenomenon was considered in \cite{Ternov2019}.  However, since the effective densities $n^{(e)}-\frac{1}{2}n^{(n)}$ in the center and in the shell are both extremely large (even if these values are of different sign), and the size of the magnetar is rather small in comparison with the neutrino spin rotation length, the adiabaticity condition can be fulfilled only for neutrinos with rather low energy. Thus, the high-energy neutrino  conserves its helicity while propagating inside the magnetar.

Since inside the magnetar the neutrino with electron flavor propagates in very dense matter, the magnetic field does not influence the neutrino propagation significantly.
So, with high accuracy we can assume that the neutrino inside the magnetar is non-oscillating and the spin rotation effect is also absent.

Since we suppose that at the edge of the magnetar the matter density decreases very fast compared to the neutrino oscillation and spin rotation typical lengths, then right after propagating in the interior of the magnetar, where the density is very high, the neutrino occurs in the region with the magnetic field only. Therefore, for neutrino outside the magnetar in Eq. \eqref{QQ} we can set $f^{(e)\mu}=0$ and $f^{(\mathrm{N})\mu}=0$. According to the Standard Model predictions, the transition magnetic and electric moments, which are taken into account in Eq. \eqref{QQ} by introducing the Hermitian matrices $\mathds{M}_h$ and $\mathds{M}_{ah}$, are much smaller than the diagonal magnetic moments. It was demonstrated in \cite{PR2020,adia_arxiv} that the transition magnetic moments cause the resonance behavior of the spin-flavor probabilities in magnetic field. However, for the resonance to take place, both high field and high neutrino energy are necessary (e.g., if $B=10^{16}$ Gauss, the neutrino energy of about $2$ GeV is required for the resonance to take place). In this paper we consider only the neutrinos with the energies, which do not reach the resonance value, and then both the electric and magnetic neutrino transition moments can be neglected. That is, the neutrino evolution in the magnetosphere is governed by the equation
\begin{equation}\label{QQ-magn}
\begin{array}{l}\displaystyle
\left(\mathrm{i} \frac{d}{d \tau}
\mathds{I}-\mathds{M}-\mu_{0} \mathds{M} \gamma^{5}
\gamma^{\mu *} F_{\mu \nu} u^{\nu} \right)
\varPsi(\tau) =0.
\end{array}
\end{equation}

Since the mass matrix $\mathds{M}$ commutes with the evolution equation, the neutrino wave function can be presented as a linear combination of the wave functions of the mass eigenstates similarly to the vacuum case, and the mixing angles for such states are equal to the vacuum mixing angles.
When the direction of the magnetic field is constant,
the spin integral of motion ${\cal S}$ exists
\begin{equation}\label{xxl171}
{\cal S}=\gamma^{5}\gamma^{\mu}{\,}^{\star\!\!}F_{\mu\nu}u^{\nu}/N,\quad
N=\displaystyle \sqrt{\displaystyle
	u_{\mu}{\,}^{\star\!\!}F^{\mu\alpha}{\,}^{\star\!\!}F_{\alpha\nu}u^{\nu}}.
\end{equation}
\noindent The operator ${\cal S}$ defines the projection of the spin on the direction of the magnetic field in the neutrino rest frame. The corresponding neutrino polarization vector is as follows
\begin{equation}
\bar{s}^\mu = - {\,}^{\star\!\!}F_{\mu\nu}u^{\nu}/N.
\end{equation}
For neutrino propagating in the purely magnetic field with the induction $\bf{B}$ we have $N=|{\bf B}| \sqrt{u_0^2 - |{\bf u}|^2\cos^2{\vartheta}}$, where $\vartheta$ is the angle between the neutrino velocity and the magnetic induction in the laboratory reference frame.

Since we have found the spin integral of motion for the neutrino mass eigenstates, the exact solutions of the evolution equation can be obtained. Using these solutions, we can derive the analytical expressions for the spin-flavor transition probabilities in three-flavor model even if the value of the field is not constant.
If the direction of the magnetic field varies slowly, then the operator ${\cal S}$ can be considered as an approximate integral of motion (see \cite{adia_arxiv}), and the expressions for the transition probabilities become approximate ones.

The formulas for the probabilities of all possible spin-flavor transitions as functions of the neutrino proper time $\tau$ are derived in Appendix.
Since the vacuum oscillation lengths are much smaller than the distance from the center of the magnetar to the neutrino detection point, only the values of the transition probabilities, averaged over the oscillation period, are of practical interest. In other words, the total transition probability should be integrated from zero to infinity over the neutrino proper time $\tau$. Here we present only the final result for the averaged probabilities of transitions from left-handed neutrino state with electron state to the left-handed and right-handed states with flavor $\beta$ ($\beta = e , \mu, \tau$).
\begin{equation}\label{7_16}
\begin{array}{l}
{{W}}_{{(e)L
	}\rightarrow{(\beta)}L}=\displaystyle
\sum\limits_{k=1}^{3}|U_{e k}|^{2}|U_{\beta
	k}|^{2}(\cos^{2}\Phi_{k} + (\bar{s} s_{sp})^2\sin^{2}\Phi_{k}),\\
{{W}}_{{(e)L}\rightarrow{(\beta)R}} =
\displaystyle \sum\limits_{k=1}^{3}|U_{e
	k}|^{2}|U_{\beta k}|^{2}(1- (\bar{s} s_{sp})^2)\sin^{2}\Phi_{k},
\end{array}
\end{equation}
\noindent where $s_{sp}^\mu$ is the polarization vector, which defines initial neutrino helicity, and
\begin{equation}\label{9z_gen}
\Phi_{k} =\frac{m_{k}\mu_{0}}{|{\bf u}|}\int\limits_{R_{0}}^{\infty}Ndr.
\end{equation}

The helicity-survival probability, which is the sum of transition probabilities to the left-handed states with all three neutrino flavors $\beta$, is given by the formula
\begin{equation}\label{SpinS}
{{W}}_{{(e)L
	}\rightarrow{}L}=\displaystyle
\sum\limits_{k=1}^{3}|U_{e k}|^{2}|(\cos^{2}\Phi_{k} + (\bar{s} s_{sp})^2\sin^{2}\Phi_{k})
\end{equation}
\noindent This probability limits the amount of neutrinos, which can possibly be detected in a terrestrial detector, since it can be assumed with high accuracy that only left-handed neutrinos interact with Standard Model particles.

The corresponding spin-flip probability
\begin{equation}
{{W}}_{{(e)L}\rightarrow{R}} =
\displaystyle \sum\limits_{k=1}^{3}|U_{e
	k}|^{2}|(1- (\bar{s} s_{sp})^2)\sin^{2}\Phi_{k}
\end{equation}
\noindent determines the difference between the initial flux of neutrinos and the flux of neutrinos of all flavors, which can be detected in terrestrial measurements, since the right-handed neutrinos are almost sterile.

Let us consider the most illustrative case of the
quasi-degenerate mass spectrum, when all the neutrino masses $m_{k}\approx m$ are approximately equal. In this case the helicity-survival probability and the spin-flip probability are as follows
\begin{equation}\label{qdeg}
\begin{array}{l}
W_{(e)L \rightarrow L} = \cos^{2}\Phi + (\bar{s} s_{sp})^2\sin^{2}\Phi,\\
W_{(e)L \rightarrow R} = \left(1 - (\bar{s} s_{sp})^2\right) \sin^{2}\Phi,
\end{array}
\end{equation}
\noindent where $\Phi\approx \Phi_k$ ($k=1,2,3$).

If we knew all the neutrino masses, the exact values of the mixing parameters and the magnetic field near the magnetar, then we would be able to obtain the exact values of probabilities using the formula \eqref{7_16}. Unfortunately, these parameters remain unknown. So, in the next section we consider the most simple case of quasi-degenerate mass spectrum assuming that the field of the magnetar decreases with distance as the magnetic dipole field.

Further for simplicity we also suppose $(\bar{s}s_{sp})=0$. In this case
\begin{equation}\label{cos}
\begin{array}{l}
W_{(e)L \rightarrow L} = \cos^{2}\Phi,\\
W_{(e)L \rightarrow R} = \sin^{2}\Phi.
\end{array}
\end{equation}
\noindent Actually, the value of $(\bar{s} s_{sp})$ depends on the angle $\vartheta$ between the direction of neutrino propagation and the magnetic induction (for detail see \cite{PR2020}), but for ultrarelativistic neutrinos it is nonzero only for a small region of angles near
$\vartheta = 0$ and $\vartheta = \pi$.

\section{Neutrinos from a magnetar with dipolar surface field}\label{sec4}

Let us consider a model of the field generated by the spherical magnetar with dipolar surface magnetic field
\begin{equation}\label{8}
{\bf{B}}={B_0}\frac{R_{0}^{3}}{r^{5}}(3{\bf r}({\bf m}{\bf r})-r^{2}{\bf m}).
\end{equation}
\noindent Here $R_{0}$ is the magnetar radius, ${\bf m}$ is a unity vector which defines the dipole orientation, ${\bf r}$  is the radius-vector from the center of the magnetar, and $B_0$ is the value of the magnetic field on the magnetar surface at the equator, i.e. where $({\bf m} {\bf r})=0$.

Assuming that all the neutrinos are produced in the central part of the magnetar, we obtain
\begin{equation}\label{9}
N/|{\bf u}|={B_0}\frac{R_{0}^{3}}{r^{3}}\sqrt{\sin^{2}\varphi
	+(3\cos^{2}\varphi+1)/|{\bf u}|^{2}}\approx {B_0}\frac{R_{0}^{3}}{r^{3}}\sin\varphi,
\end{equation}
\noindent where  $\varphi$  is the angle between the dipole axis and the neutrino emission
direction. Note, that if the neutrino is produced precisely in the center of the magnetar, these solutions are exact ones.

So,
\begin{equation}\label{9z}
\Phi =\frac{m\mu_{0}}{|{\bf u}|}\int\limits_{R_{0}}^{\infty}Ndr=
\frac{R_{0}}{2}m\mu_{0}{B_0}\sin\varphi=\frac{\pi{R_{0}}}{L_{0}}\sin\varphi,
\end{equation}
\noindent where
\begin{equation}\label{9zz}
{L_{0}}=\frac{2\pi}{m\mu_{0}{B_0}},
\end{equation}
\noindent is the typical spin rotation length. If ${L_{0}}$ is significantly greater than the
magnetar radius, then the effect of the magnetic field on the neutrino propagation is negligible. However, if ${L_{0}}\lesssim R_{0}$, then the effect of the magnetic field on neutrino evolution becomes non-trivial.

\begin{figure}[tbp]
	\begin{minipage}{0.45\textwidth}
		\includegraphics[width=\textwidth]{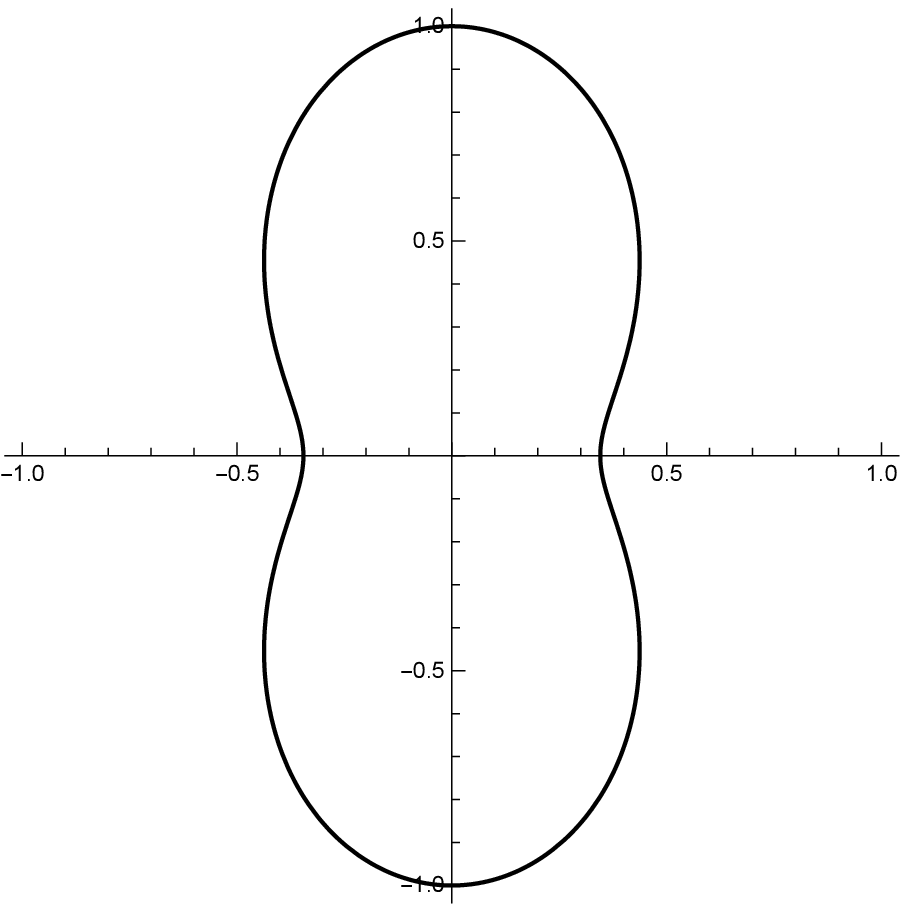}
		\caption{\label{F1}	 The angular distribution of the spin survival probability for $\kappa\approx 0.3$}
	\end{minipage}
\hfill
	\begin{minipage}{0.45\textwidth}
		\includegraphics[width=\textwidth]{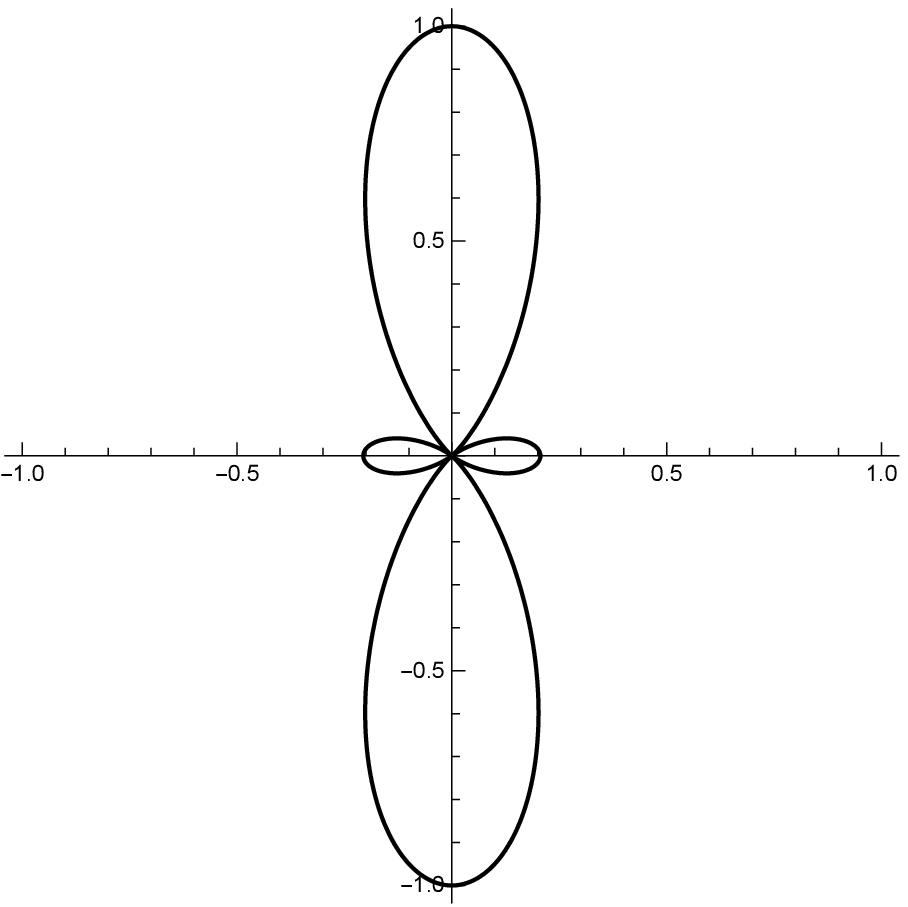}
		\caption{\label{F2} The angular distribution of the spin survival probability for $\kappa\approx 0.65$}
	\end{minipage}
\end{figure}

\begin{figure}[tbp]
	\begin{minipage}{0.45\textwidth}
		\includegraphics[width=\textwidth]{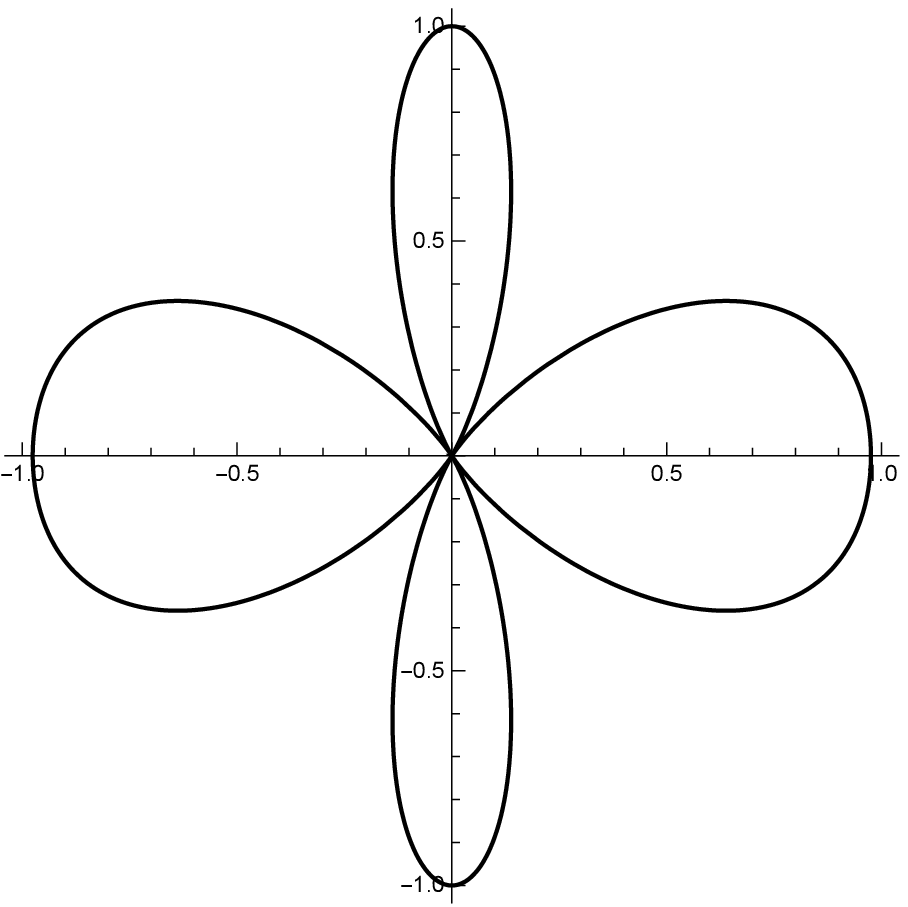}
		\caption{\label{F3} The angular distribution of the spin survival probability for $\kappa\approx 0.95$}
	\end{minipage}
\hfill
	\begin{minipage}{0.45\textwidth}
		\includegraphics[width=\textwidth]{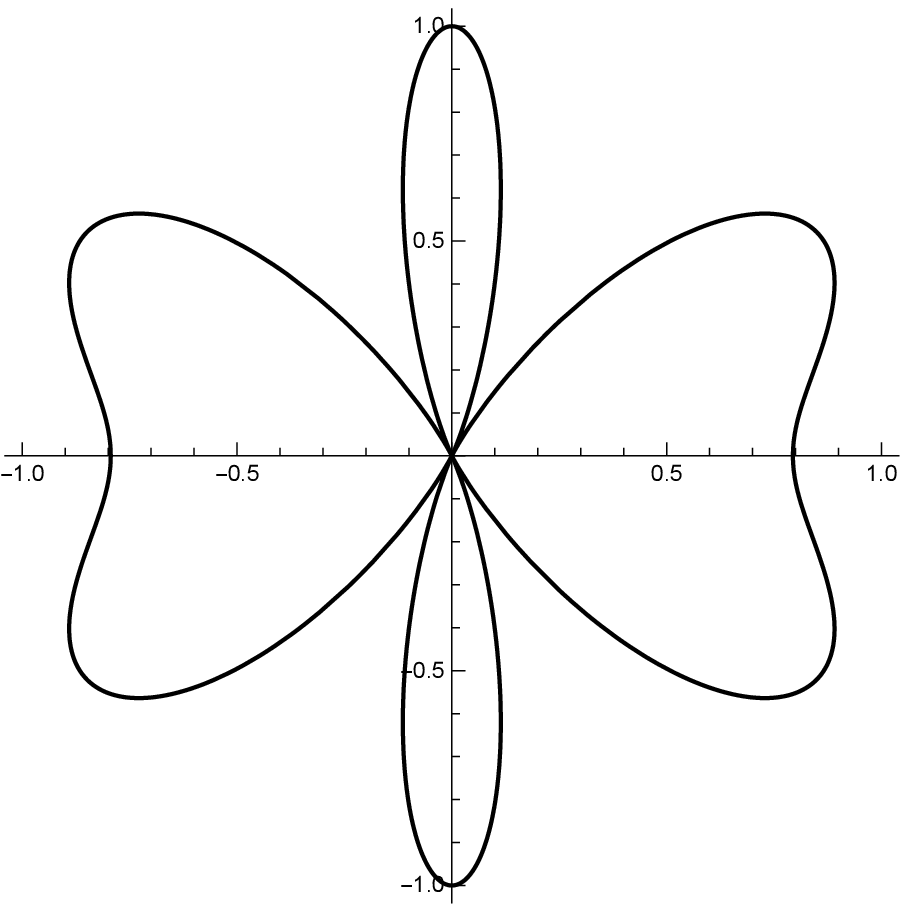}
		\caption{\label{F4} The angular distribution of the spin survival probability for $\kappa\approx 1.15$}
	\end{minipage}
\end{figure}

The helicity-survival probability Eq. \eqref{cos} in this case is given by the relation
\begin{equation}
W_{eL\rightarrow L} = \displaystyle \cos^2 \bigg(\frac{\pi R_0}{L_0}\sin\varphi\bigg) = \cos^2 \bigg(\pi \kappa\sin\varphi\bigg)
\end{equation}	
\noindent and effectively depends only on the parameter
\begin{equation}
\kappa =  R_0/L_0 = \frac{R_0}{2\pi} m \mu_0 B_0,
\end{equation}
\noindent which is proportional both to the magnetar size and the value of the surface magnetic field.

For different values of the parameter $\kappa$ the distribution of neutrino fluxes in space changes significantly. In Fig. \ref{F1} the dependence of the survival probability on the angle between the magnetic axis and the neutrino velocity is given for the left-handed neutrino with electron flavor for $\kappa = 0.3$  in polar coordinates. With the increase of the magnetar radius the anisotropy of the neutrino flux becomes more significant (see Fig. \ref{F2}--\ref{F4}). When the size of the magnetar becomes large enough (i.e. $\kappa > 1/2$), some directions appear, for which the spin-survival probability is zero. In these directions only right-handed neutrinos can reach the detector.

\section{Conclusion}

In this paper we calculate the exact spin-flavor transition probabilities for neutrino in non-homogeneous magnetic field taking into account diagonal magnetic moments only. Using the solutions of this auxiliary problem, we study neutrino propagation from a magnetar. For this purpose we make some assumptions, which in our opinion are not crucial.

Although there is a narrow region inside the magnetar, where the spin rotation effect can occur, we  assume that the neutrino conserves both its flavor and helicity inside the object. Actually, a fraction of left-handed neutrinos of other flavors in the flux at the edge of the magnetar will lead to the change of the mixing matrix parameters $U_{e k}$ in the final formulas to their effective values, which define a superposition of the mass states outside the magnetar.

Even in the region of the star, where the neutrino spin-flip transitions are possible, they are suppressed by the potentials of neutrino interaction with matter. A small fraction of right-handed particles in the flux at the edge of the magnetar will lead to the deviation of  $(\bar{s} s_{sp})$ from zero in Eq. \eqref{7_16}. Then the minimal value of the spin-survival probabilities (see Figs. \ref{F1}---\ref{F4}) will never be equal to zero.

The approximation of the neutrino mass spectrum as quasi-degenerate is made for simplicity. For large values of $\kappa$, which although seem rather non-realistic, the effect of the anisotropy of spin-survival transition probabilities will obviously vanish for real neutrinos with non-equal masses. For small $\kappa$ taking into account different neutrino masses will lead to a more complicated structure of the angular distribution of the transition probabilities. Taking into account different neutrino masses also leads to non-zero spin survival probability in the direction, where this probability takes its minimal value (compare Eq. \eqref{cos} and Eq. \eqref{7_16}). That is, the deviations from all our assumptions on neutrino propagation, in the general case result in the blurring of the central minimum of the angular distribution of the left-handed neutrino flux.

As the main qualitative result of our paper we obtain that the anisotropy in neutrino fluxes can be caused by neutrino spin-flip effect. This means that the observed neutrino activity of magnetars with extra-high magnetic fields may significantly depend on the direction of the magnetar axis. The actual magnetar neutrino flux in this case may be much greater than the flux, that is observed in terrestrial conditions

\appendix

\section{Spin-flavor transition probabilities}
	
The solutions of the quasi-classical evolution equation \eqref{QQ} can be presented with the help of the resolvent $U(\tau)$
\vspace{-6pt}
\begin{equation}\label{sf1}
\varPsi(\tau)=\frac{1}{\sqrt{2 u_0}}\, U(\tau) \varPsi_{0},
\end{equation}
\noindent where $\varPsi_{0}$ is a constant $12$-component object, which defines the initial state of the neutrino. For a neutrino pure state with a definite initial polarization it can be presented in the form
\begin{equation}\label{sf2}
\varPsi_0=\frac{1}{2}(1-\gamma^{5}\gamma_{\mu}{s}_{0}^{\mu}
) (\gamma_\mu u^\mu+1) \left(\psi^0\otimes e_{j}\right), \quad
\bar{\varPsi}_0\varPsi_0= 2.
\end{equation}
\noindent Here $\psi^0$ is a constant bispinor, $e_{j}$ is an arbitrary unit vector in the three-dimensional vector space over
the field of complex numbers, and ${s}_{0}^{\mu}$ is a $4$-vector of neutrino polarization, which satisfies the relation $(u{s}_{0})=0$.
In the general case the resolvent can be written as the multiplicative integral of Volterra (see, e.g., \cite{gantmakher}). In fact, this integral can be presented as a T-exponential.

For arbitrary external conditions the resolvent obviously can not be presented as an analytical expression.
In this paper we do not take into account neutrino transition moments, and therefore any state of neutrino in magnetic field can be considered as a superposition of neutrino mass eigenstates with the mixing angles equal to their vacuum values.
If in this case the spin integral of motion $\mathcal{S}$ exists, the analytical expression for the resolvent can be obtained
\begin{equation}\label{sol11}\displaystyle
U(\tau) = \frac{1}{2} \sum\limits_{i=1,2,3} \sum\limits_{\zeta=\pm 1}e^{-\ii m_i
	\int\limits_0^\tau (1-\zeta \mu_0 N(\tilde{\tau})) d\tilde{\tau}}  \left(
{1-\zeta\gamma^5 \gamma^\mu \bar{s}_\mu} \right)\mathds{P}^{(i)},
\end{equation}
\noindent where $\mathds{P}^{(i)}$ is the projection operator on the the state with the mass $m_i$.

To describe the neutrino pure state, which is characterized by the definite initial flavor $\alpha$ and definite initial helicity $\zeta_\alpha$ ($\zeta_\alpha = - 1$ for left-handed neutrinos and $\zeta_\alpha = 1$ for right-handed neutrinos)  we use the spin-flavor density matrices
\begin{equation}\label{rho4}
\rho_{(\alpha), \zeta_\alpha}(\tau)=\frac{1}{4u^{0}}U(\tau)\big(\gamma^{\mu}u_{\mu}+1\big)\left(1-
\zeta_\alpha
\gamma^{5}\gamma_\mu {s}^{\mu}_{sp}
\right){\mathds
	P}_{0}^{(\alpha)}\bar{U}(\tau),
\end{equation}
\noindent where $s_{sp}^\mu = \{|{\bf u}|, u^0 {\bf u}/ {|{\bf u}|}\}$. In Eq. \eqref{rho4} the projection operator ${\mathds
	P}_{0}^{(\alpha)}$ determines the neutrino initial flavor state, and the projection operator
$(1-\zeta_\alpha \gamma^5 \gamma_\mu s_{sp}^\mu)/2$  determines the neutrino initial helicity state.

In accordance with  the fundamental principles	of quantum mechanics the probability of transition from the state with $\alpha$ and $\zeta_\alpha$ to the state with
$\beta$ and $\zeta_\beta$ during proper time $\tau$ is
\begin{equation}\label{ver}
W_{(\alpha),\zeta_\alpha \rightarrow(\beta) ,\zeta_\beta}=\Tr\left\{ \rho_{(\alpha),\zeta_\alpha}(\tau)\rho_{(\beta),\zeta_\beta}^{\dag}
(\tau=0)\right\}.
\end{equation}
\noindent Taking into account the relation
\begin{equation}\label{tr1}
{\mathrm{Tr}}\left\{\mathds{P}^{(k)}{\mathds{U}}^{\dag}\mathds{P}^{(\alpha)}{\mathds{U}}
\mathds{P}^{(l)}{\mathds{U}}^{\dag}\mathds{P}^{(\beta)}{\mathds{U}}\right\}=U_{\alpha
	k}^{*}U_{\alpha l}U_{\beta l}^{*}U_{\beta
	k}.
\end{equation}
\noindent we obtain the spin-flavor transition probabilities
\begin{multline}\label{36x}
{W}_{{(\alpha),\zeta_\alpha}\rightarrow{(\beta), \zeta_\beta}}=\frac{1}{4}
\sum\limits_{k,l=1}^{3}U_{\alpha k}^{*}U_{\alpha l}U_{\beta l}^{*}U_{\beta
	k}\sum\limits_{\zeta,\zeta'=\pm 1} e^{-\ii (m_{k}\tau-\zeta\Phi_{k}(\tau))} e^{
	\,\ii (m_{l}\tau-\zeta'\Phi_{l}(\tau))}\\
\times\left\{\frac{1+\zeta_{\alpha}\zeta_{\beta}}{2}
\big(1-\zeta_{\alpha}(\zeta+\zeta')(\bar{s}s_{sp})+\zeta\zeta'(\bar{s}s_{sp})^{2}\big)
+\frac{1-\zeta_{\alpha}\zeta_{\beta}}{2}
\zeta\zeta'\left(1-(\bar{s}s_{sp})^{2}\right)\right\},
\end{multline}
\noindent where
\begin{equation}\label{1x}
\Phi_{k}(\tau)=m_{k}\mu_{0}\int\limits_{0}^{\tau}Nd\tau.
\end{equation}

The probability of the transition to the state with the same helicity as the initial one is as follows
\begin{equation}\label{3}
\begin{array}{l}
W_{(\alpha),\zeta_\alpha\rightarrow (\beta),\zeta_\beta =\zeta_\alpha}=\sum\limits_{k=1}^{3}|U_{\alpha k}|^{2}|U_{\beta
	k}|^{2}\left(\cos^{2}\Phi_{k}(\tau)
+(\bar{s}s_{sp})^{2}\sin^{2}\Phi_{k}(\tau)\right)\\
\displaystyle +\sum\limits_{k>l}^{3} R_{\alpha\beta
	kl}\Big\{\frac{1}{2}\big(1-\zeta_{\alpha}(\bar{s}s_{sp})\big)^{2}\cos\big((m_{k}-m_{l})\tau-
\Phi_{k}(\tau)+\Phi_{l}(\tau)\big)\\ \phantom{\sum\limits_{k>l}^{3} R_{\alpha\beta
		kl}z,}+\displaystyle \frac{1}{2}
\big(1+\zeta_{\alpha}(\bar{s}s_{sp})\big)^{2}\cos\big((m_{k}-m_{l})\tau+
\Phi_{k}(\tau)-\Phi_{l}(\tau)\big) \\  \phantom{\sum\limits_{k>l}^{3} R_{\alpha\beta
		kl}z,}+
\big(1-(\bar{s}s_{sp})^{2}\big)\cos(m_{k}-m_{l})\tau
\cos\big(\Phi_{k}(\tau)+\Phi_{l}(\tau)\big)\Big\}\\
\displaystyle +\sum\limits_{k>l}^{3} I_{\alpha\beta
	kl}\Big\{\frac{1}{2}\big(1-\zeta_{\alpha}(\bar{s}s_{sp})\big)^{2}\sin\big((m_{k}-m_{l})\tau-
\Phi_{k}(\tau)+\Phi_{l}(\tau)\big) \\ \phantom{\sum\limits_{k>l}^{3} I_{\alpha\beta
		kl}z,}+\displaystyle \frac{1}{2}
\big(1+\zeta_{\alpha}(\bar{s}s_{sp})\big)^{2}\sin((m_{k}-m_{l})\tau+
\Phi_{k}(\tau)-\Phi_{l}(\tau))\\  \phantom{\sum\limits_{k>l}^{3} I_{\alpha\beta
		kl}z,}+
\big(1-(\bar{s}s_{sp})^{2}\big)\sin(m_{k}-m_{l})\tau
\cos\big(\Phi_{k}(\tau)+\Phi_{l}(\tau)\big)\Big\}.
\end{array}
\end{equation}
\noindent	The corresponding spin-flip probability is given by the expression
\begin{equation}\label{4x}
\begin{array}{l}
W_{(\alpha),\zeta_\alpha\rightarrow (\beta),\zeta_\beta\neq\zeta_\alpha}=\sum\limits_{k=1}^{3}|U_{\alpha k}|^{2}|U_{\beta
	k}|^{2}\big(1- (\bar{s}s_{sp})^{2}\big)\sin^{2}\Phi_{k}(\tau)\\
\displaystyle +\sum\limits_{k>l}^{3} R_{\alpha\beta
	kl}\Big\{\frac{1}{2}\big(1-(\bar{s}s_{sp})^{2}\big)\cos\big((m_{k}-m_{l})\tau-
\Phi_{k}(\tau)+\Phi_{l}(\tau)\big) \\ \phantom{\sum\limits_{k>l}^{3} R_{\alpha\beta
		kl}z,}+\displaystyle \frac{1}{2}
\big(1-(\bar{s}s_{sp})^{2}\big)\cos\big((m_{k}-m_{l})\tau
+\Phi_{k}(\tau)-\Phi_{l}(\tau)\big) \\  \phantom{\sum\limits_{k>l}^{3}
	R_{\alpha\beta kl}z,}-
\big(1-(\bar{s}s_{sp})^{2}\big)\cos(m_{k}-m_{l})\tau
\cos\big(\Phi_{k}(\tau)+\Phi_{l}(\tau)\big)\Big\}\\
\displaystyle +\sum\limits_{k>l}^{3} I_{\alpha\beta
	kl}\Big\{\frac{1}{2}\big(1-(\bar{s}s_{sp})^{2}\big)\sin\big((m_{k}-m_{l})\tau
-\Phi_{k}(\tau)+\Phi_{l}(\tau)\big) \\ \phantom{\sum\limits_{k>l}^{3} I_{\alpha\beta
		kl}z,}+\displaystyle \frac{1}{2}
\big(1-(\bar{s}s_{sp})^{2}\big)\sin\big((m_{k}-m_{l})\tau
+\Phi_{k}(\tau)-\Phi_{l}(\tau)\big) \\  \phantom{\sum\limits_{k>l}^{3}
	I_{\alpha\beta kl}z,}-
\big(1-(\bar{s}s_{sp})^{2}\big)\sin(m_{k}-m_{l})\tau
\cos\big(\Phi_{k}(\tau)+\Phi_{l}(\tau)\big)\Big\}.
\end{array}
\end{equation}
\noindent In the formulas \eqref{3}, \eqref{4x} the following notations are used
\begin{equation}\label{2}
R_{\alpha\beta kl}\equiv {\mathrm{Re}}\,
U_{\alpha k}^{*}U_{\alpha l}U_{\beta l}^{*}U_{\beta k},\quad I_{\alpha\beta
	kl}\equiv {\mathrm{Im}}\,
U_{\alpha k}^{*}U_{\alpha l}U_{\beta l}^{*}U_{\beta k}.
\end{equation}

The obtained formulas coincide with the standard expressions for neutrino
oscillations in vacuum (see, e.g.,\cite{Giunti_book}), if we neglect the effect of
the field, i.e. assume $N=0$.
If we consider the two-flavor model, which takes into account the effect of the
magnetic field, the formula  \eqref{2} coincides with the expression obtained
in \cite{PR2020} (see also \cite{Dvornikov2007,Popov2019}).

Since the distance between the neutrino source and the detector is much greater than the typical oscillation length, only the averaged over the oscillation period probabilities can be of practical use
\begin{equation}
\begin{array}{l}
W_{(\alpha),\zeta_\alpha\rightarrow (\beta), \zeta_\beta = \zeta_\alpha}=\sum\limits_{k=1}^{3}|U_{\alpha k}|^{2}|U_{\beta
	k}|^{2}\left(\cos^{2}\Phi_{k}(\tau)
+(\bar{s}s_{sp})^{2}\sin^{2}\Phi_{k}(\tau)\right)\\
W_{(\alpha),\zeta_\alpha\rightarrow (\beta), \zeta_\beta\neq\zeta_\alpha}=\sum\limits_{k=1}^{3}|U_{\alpha k}|^{2}|U_{\beta
	k}|^{2}\big(1- (\bar{s}s_{sp})^{2}\big)\sin^{2}\Phi_{k}(\tau)
\end{array}
\end{equation}

\acknowledgments
{The authors are grateful to A.V. Borisov, E.M. Murchikova,  D.D. Sokolov, V.A. Sokolov, I.P. Volobuev and V.Ch. Zhukovsky for fruitful discussions. A.V.C. acknowledges support from the Foundation
for the advancement of theoretical physics and mathematics
``BASIS'' (Grant No. 19-2-6-100-1).}

\end{document}